\documentclass{article}
\usepackage{spconf,amsmath,graphicx}

\usepackage{enumitem}
\setlist{nosep, leftmargin=14pt}

\usepackage{mwe} % to get dummy images

\title{Efficient Knowledge Distillation of SAM for Medical Image Segmentation}
\name{Kunal Dasharath Patil$^1$, Gowthamaan Palani$^1$, Ganapathy Krishnamurthi$^1$}
\address{$^1$Indian Institute of Technology, Madras, India}
\begin{document}
%\ninept
%
\maketitle
\begin{abstract}
  The Segment Anything Model (SAM) has set a new standard in interactive image segmentation, offering robust performance across various tasks. However, its significant computational requirements limit its deployment in real-time or resource-constrained environments. To address these challenges, we propose a novel knowledge distillation approach, KD SAM, which incorporates both encoder and decoder optimization through a combination of Mean Squared Error (MSE) and Perceptual Loss. This dual-loss framework captures structural and semantic features, enabling the student model to maintain high segmentation accuracy while reducing computational complexity. Based on the model evaluation on datasets, including Kvasir-SEG, ISIC 2017, Fetal Head Ultrasound, and Breast Ultrasound, we demonstrate that KD SAM achieves comparable or superior performance to the baseline models, with significantly fewer parameters. KD SAM effectively balances segmentation accuracy and computational efficiency, making it well-suited for real-time medical image segmentation applications in resource-constrained environments.
\end{abstract}
\section{Introduction}
\label{sec:intro}

Interactive image segmentation has become a cornerstone in numerous applications, including medical imaging, autonomous driving, and augmented reality. The Segment Anything Model (SAM) \cite{kirillov2023segment} has established itself as a powerful tool in this domain, leveraging a Vision Transformer (ViT) \cite{dosovitskiy2020image} encoder and prompt-guided mask decoder to achieve high segmentation accuracy across diverse datasets. However, the significant computational demands of SAM hinder its deployment in real-time and resource-constrained environments, such as mobile devices and edge platforms.

MobileSAM \cite{zhang2023faster} addresses these limitations by replacing the ViT encoder with ViT-Tiny, significantly reducing the model size and inference time while maintaining competitive performance. Despite these advances, the segmentation quality, particularly for complex tasks such as medical imaging, is compromised due to the reduced capacity of the ViT-Tiny encoder.

In this work, we propose a novel decoupled knowledge distillation approach that enhances both the encoder and decoder components. By incorporating both Mean Squared Error (MSE) and perceptual loss \cite{johnson2016perceptual}, our method captures structural and semantic features, resulting in a robust model capable of high-precision segmentation with reduced computational costs. This approach addresses MobileSAM's challenges, making our method more effective for medical imaging tasks where accuracy and efficiency are crucial.

\section{Related Work}
\label{sec:format}

Knowledge distillation is a powerful technique to transfer knowledge from a large, complex model to a smaller model while maintaining considerable performance. Initially developed for classification tasks \cite{guan2020differentiable}, it has been adapted for dense prediction tasks such as semantic segmentation \cite{liu2019structured} and object detection \cite{chen2017learning}. The goal is to align the student model feature representations with the teacher, preserving spatial and semantic information. Strategies like pixel-wise feature matching and channel-wise correlations have been employed to ensure the student model replicates the teacher's performance with reduced computational cost.\\
The Segment Anything Model (SAM) \cite{kirillov2023segment} represents a significant advancement in interactive segmentation, leveraging a powerful ViT-H encoder and prompt-guided mask decoder to handle diverse segmentation tasks. However, its high computational demands have limited its real-world applications. Several variants of SAM have been developed to address these limitations. MobileSAM \cite{zhang2023faster} replaces the ViT-H encoder with ViT-Tiny, reducing the model size and inference time while maintaining competitive performance. However, MobileSAM struggles with fine-grained segmentation tasks, particularly in medical imaging.\\
On the other hand, FastSAM \cite{zhao2023fast} employs a YOLACT-based \cite{bolya2019yolact} instance segmentation model combined with heuristic post-processing rules for object selection to achieve faster segmentation. However, this comes at the cost of segmentation quality, as it diverges from SAM’s interactive segmentation principles, making it less suitable for high-precision tasks. EfficientSAM \cite{xiong2024efficientsam} focuses on improving training efficiency using masked image pre-training, but it does not significantly improve inference speed over MobileSAM. EdgeSAM \cite{zhou2023edgesam}, which also employs encoder-only distillation inspired by MobileSAM, struggled to bridge the performance gap despite utilizing various efficient backbones and extended training schedules. MSE loss alone was insufficient to align the student and teacher model features, particularly in complex segmentation scenarios. While additional task-specific information using prompt-based distillation provided some improvements, the limitations of relying solely on MSE persisted.\\
In response to these challenges, our work implements a decoupled knowledge distillation framework, optimizing the encoder and decoder separately. By incorporating MSE and perceptual loss, we ensure better alignment of the student feature representations with the teacher, capturing both low-level structural details and high-level semantic information. This approach not only preserves the segmentation accuracy of SAM but also significantly reduces the computational burden, making it suitable for real-time applications, particularly in resource-constrained environments like mobile devices and edge computing platforms.

\section{Method}
\label{sec:pagestyle}

The proposed method adapts the Segment Anything Model (SAM) for medical image segmentation through a tailored decoupled knowledge distillation process as done by MobileSAM. This approach addresses the computational limitations of SAM's Vision Transformer (ViT) encoder by distilling its knowledge into a lightweight ResNet \cite{he2016deep} based encoder. As shown in Figure 1, the methodology involves a two-phase process: the first phase focuses on encoder distillation, and the second phase involves fine-tuning the decoder. This decoupled strategy allows efficient training with limited computational resources while ensuring high segmentation accuracy on medical datasets.

\begin{figure}[htb]
  \centerline{\includegraphics[width=8.5cm]{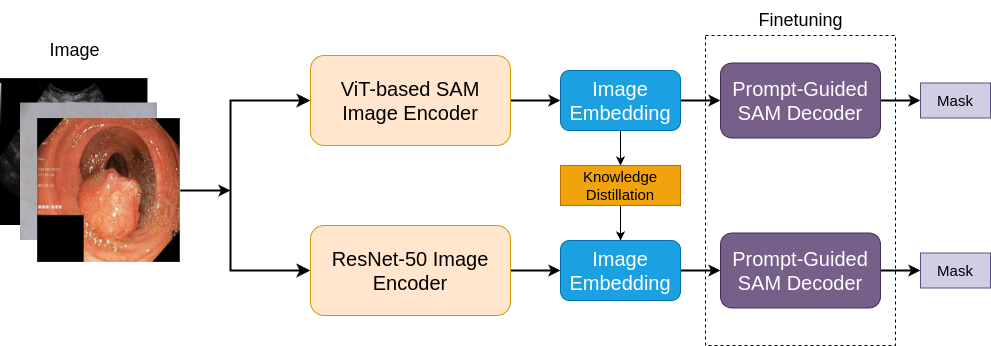}}
  %  \vspace{2.0cm}
  \centerline{}
  \caption{Knowledge Distillation Process for the KD SAM Model}
\end{figure}

\subsection{Encoder Knowledge Distillation}
The first phase of the methodology involves distilling the knowledge from SAM's ViT encoder to a more computationally efficient ResNet-50 encoder. The selection of ResNet-50 as several considerations drove the student model. First, ResNet-50, with its deep residual learning framework, effectively mitigates the vanishing gradient problem, enabling the model to learn deep feature representations without degradation in performance. This property is particularly beneficial when attempting to capture the complex structures inherent in medical images. Second, ResNet-50 strikes an optimal balance between model size and performance, making it suitable for deployment in resource-constrained environments, such as mobile devices or edge computing platforms. Its significantly lower parameter count than the ViT encoder reduces computational requirements, facilitating real-time inference without compromising segmentation accuracy.

The distillation process employs a combined loss function that integrates Mean Squared Error (MSE) and Perceptual loss. MSE measures the pixel-wise differences between the feature maps of the teacher model (ViT encoder) and the student model (ResNet-50 encoder), ensuring the essential structural information captured by the high-dimensional feature space of the ViT model is transferred effectively to the ResNet model. However, relying solely on MSE loss can lead to a loss of perceptual quality in the distilled features, particularly for fine-grained details critical in medical image segmentation. Perceptual loss is incorporated to address this, which leverages pre-trained feature extractors, such as VGG \cite{simonyan2014very} networks, to capture high-level semantic similarities between the two models. This loss function evaluates the distance between the feature representations of the teacher and student models at multiple layers, ensuring that the student model not only replicates the low-level details but also learns the perceptual features necessary for distinguishing complex structures. This dual loss approach ensures that the ResNet-50 encoder can approximate the performance of the ViT encoder while being computationally efficient, making it a practical solution for medical image segmentation tasks.

The combined loss function $L_{Combined}$ is defined as:
\begin{equation}
  L_{Combined} = L_{MSE} + L_{P}
\end{equation}
Where, $L_{MSE}$ represent MSE Loss and $ L_{P}$ is Perceptual loss. The MSE loss, given as
\begin{equation}
  L_{MSE} = \frac{1}{N} \sum_{i=1}^{N} \left( f^{T}(x_i) - f^{S}(x_i) \right)^2
\end{equation}
where $f^{T}(x_i)$ and $f^{S}(x_i)$ are the feature maps from the teacher and student encoders, respectively. The Perceptual loss $L_{P}$ is calculated using feature activations from selected layers $l$ of a pre-trained VGG network:
\begin{equation}
  L_{P} = \sum_{l} \frac{1}{C_l H_l W_l} \sum_{c=1}^{C_l} \sum_{h=1}^{H_l} \sum_{w=1}^{W_l} \left( \phi_l^{T}(x_i) - \phi_l^{S}(x_i) \right)^2
\end{equation}
In this formula, $\phi_l^{T}(x_i)$ and $\phi_l^{S}(x_i)$ denote the feature maps from layer $l$ of the ViT and ResNet encoders, respectively. The terms $C_l, H_l$ and  $W_l$ represent the number of channels, height, and width of the feature map at layer $l$. The perceptual loss ensures that the student encoder captures the semantic content and overall structure of the input images, which is crucial for complex medical image segmentation tasks.

\subsection{Decoder Fine-Tuning}
Unlike traditional distillation methods, which train the encoder and decoder concurrently, we employ a decoupled approach. Coupled distillation optimizes the student encoder and decoder, allowing the decoder to adapt to any changes in the feature representations generated by the student encoder. This joint optimization ensures that the entire student network (encoder and decoder) aligns closely with the teacher model. However, this approach is computationally expensive and challenging to implement on resource-constrained devices.

Our decoupled distillation method trains the encoder independently to learn the feature embeddings of the ViT-based encoder from SAM. This training strategy is essential to reduce the computational burden and allows us to focus on distilling the encoder first without the complexities of optimizing the entire network. However, this independence means that the feature embeddings generated by the distilled encoder may not be fully compatible with the pre-trained SAM decoder. As a result, the decoder needs to be fine-tuned to align with the distilled encoder.

To achieve this, we fine-tune the decoder using a Dice Loss. The Dice Loss is particularly effective for medical image segmentation because it maximises the overlap between the predicted segmentation mask and the ground truth, handling the class imbalance common in medical datasets. The Dice Loss is defined as:
\begin{equation}
  L_{Dice} = 1 - \frac{2 \sum_{i=1}^{N} p_i g_i}{\sum_{i=1}^{N} p_i + \sum_{i=1}^{N} g_i}
\end{equation}
Where $p_i$ represents the predicted segmentation mask, $g_i$ represents the ground truth mask, and $N$ is the total number of pixels.
During fine-tuning, the encoder's weights are kept frozen to retain the distilled knowledge while the decoder is trained on the dataset. This selective training approach is computationally efficient and ensures that the decoder adapts to the unique features generated by the distilled encoder, achieving high segmentation accuracy.

\section{Experimental Setup}
\label{sec:typestyle}

The training process for the knowledge distillation framework was carried out on multiple medical imaging datasets, including Kvasir-SEG \cite{jha2020kvasir}, ISIC 2017 \cite{cassidy2022analysis}, Fetal Head Ultrasound \cite{van2018automated}, and Breast Ultrasound \cite{al2020dataset}. These datasets were chosen for their diversity and relevance to the segmentation tasks, providing a comprehensive evaluation of the model's ability to generalize across various medical image types. Each dataset represents unique challenges, from polyp segmentation in gastrointestinal images to the precise delineation of anatomical structures in fetal head ultrasound scans.\\
The training involved two key phases: encoder distillation and decoder fine-tuning. The encoder distillation phase trains the ResNet-50 student model to learn the SAM ViT-H encoder representations. The ResNet-50 architecture was modified to reduce the channel dimensions from 2048 to 256 and incorporate upsampling layers to match the spatial dimensions of the teacher model’s outputs.\\
For Encoder distillation, the number of epochs is set to 100 with a batch size of 16. Early stopping was applied to halt training when the validation loss failed to improve, ensuring the model stopped once optimal performance was reached without overfitting. The Adam optimizer \cite{kingma2014adam} with a learning rate of 0.0001 and weight decay of 0.001 was used, and a learning rate scheduler reduced the learning rate by a factor of 0.1 if the validation loss plateaued. In decoder fine-tuning, the SAM decoder coupled with the distilled encoder was fine-tuned using Dice Loss, which maximizes the overlap between predicted segmentation masks and ground truth labels, making it particularly effective for medical imaging tasks.

\section{Results}
\label{sec:majhead}

The performance of the KD SAM model was evaluated on a separate test dataset using the Dice Coefficient metric across four medical imaging datasets: Kvasir-SEG, ISIC 2017, Fetal Head Ultrasound, and Breast Ultrasound, and compared against the baseline models, SAM and MobileSAM. As shown in Table 1, the results demonstrate that KD SAM achieves comparable or superior performance to the baseline models across most datasets. KD SAM maintains high segmentation accuracy, with Dice Coefficients close to or exceeding those of both SAM and MobileSAM.

\begin{table}[ht]
  \centering
  \caption{Dice Score comparison of KD SAM with baseline models across different medical imaging datasets.}
  \resizebox{\columnwidth}{!}{%
    \begin{tabular}{|l|c|c|c|}
      \hline
      \textbf{Dataset}  & \textbf{SAM} & \textbf{MobileSAM} & \textbf{KD SAM} \\ \hline
      Kvasir-SEG        & 0.8715       & 0.8719             & 0.8586          \\ \hline
      Fetal Head        & 0.9755       & 0.9734             & 0.9774          \\ \hline
      ISIC 2017         & 0.9091       & 0.9055             & 0.9114          \\ \hline
      Breast Ultrasound & 0.9051       & 0.8985             & 0.8216          \\ \hline
    \end{tabular}
  } % End of resizebox
  \label{table:quantitative_results}
\end{table}

The parameter comparison further underscores the model's efficiency, where KD SAM reduces the parameter count to 26.4 million compared to 632 million for SAM. While MobileSAM uses fewer parameters at 5 million, KD SAM strikes a better balance between model complexity and segmentation accuracy, particularly in medical imaging scenarios where precision is crucial.

\begin{figure}[htb]

  \begin{minipage}[b]{1.0\linewidth}
    \centering
    \centerline{\includegraphics[width=8.5cm]{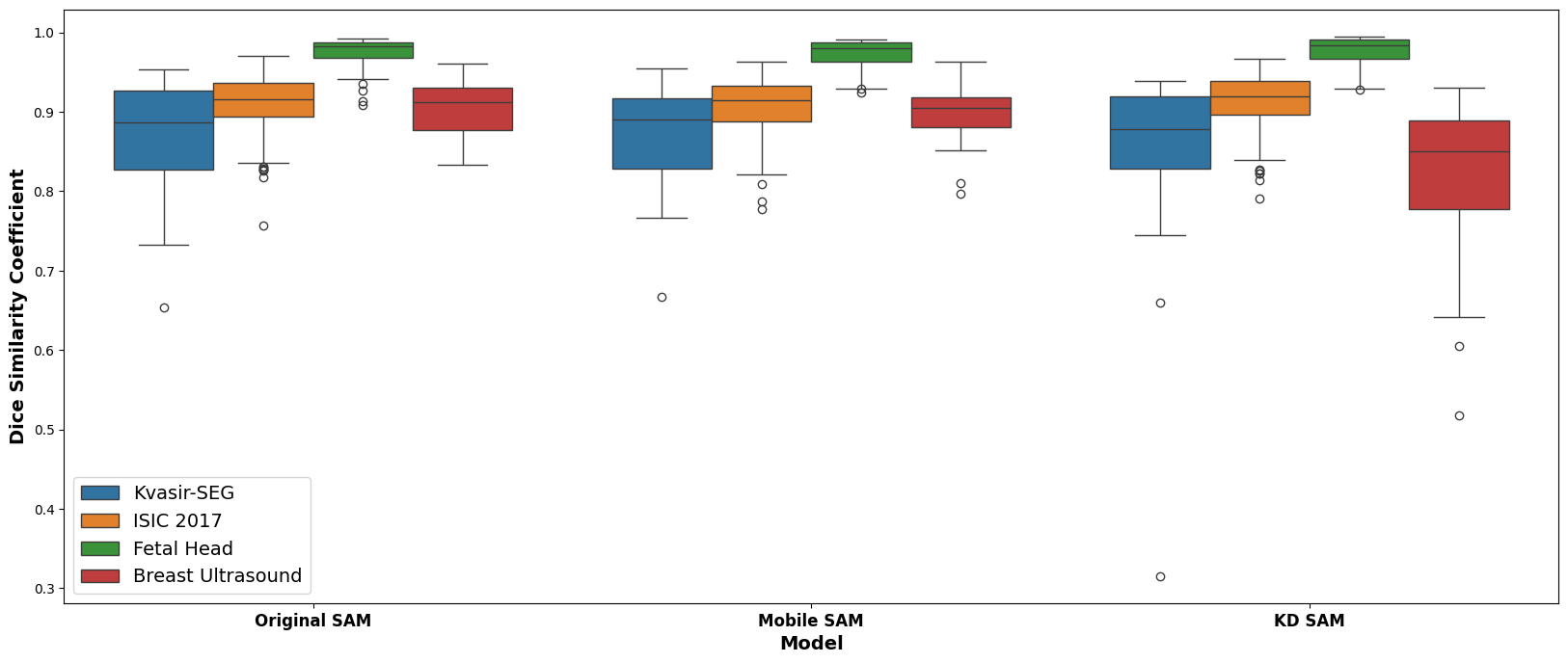}}
    %  \vspace{2.0cm}
    \centerline{(a) Box Plot}\medskip
  \end{minipage}
  \begin{minipage}[b]{1.0\linewidth}
    \centering
    \centerline{\includegraphics[width=8.5cm]{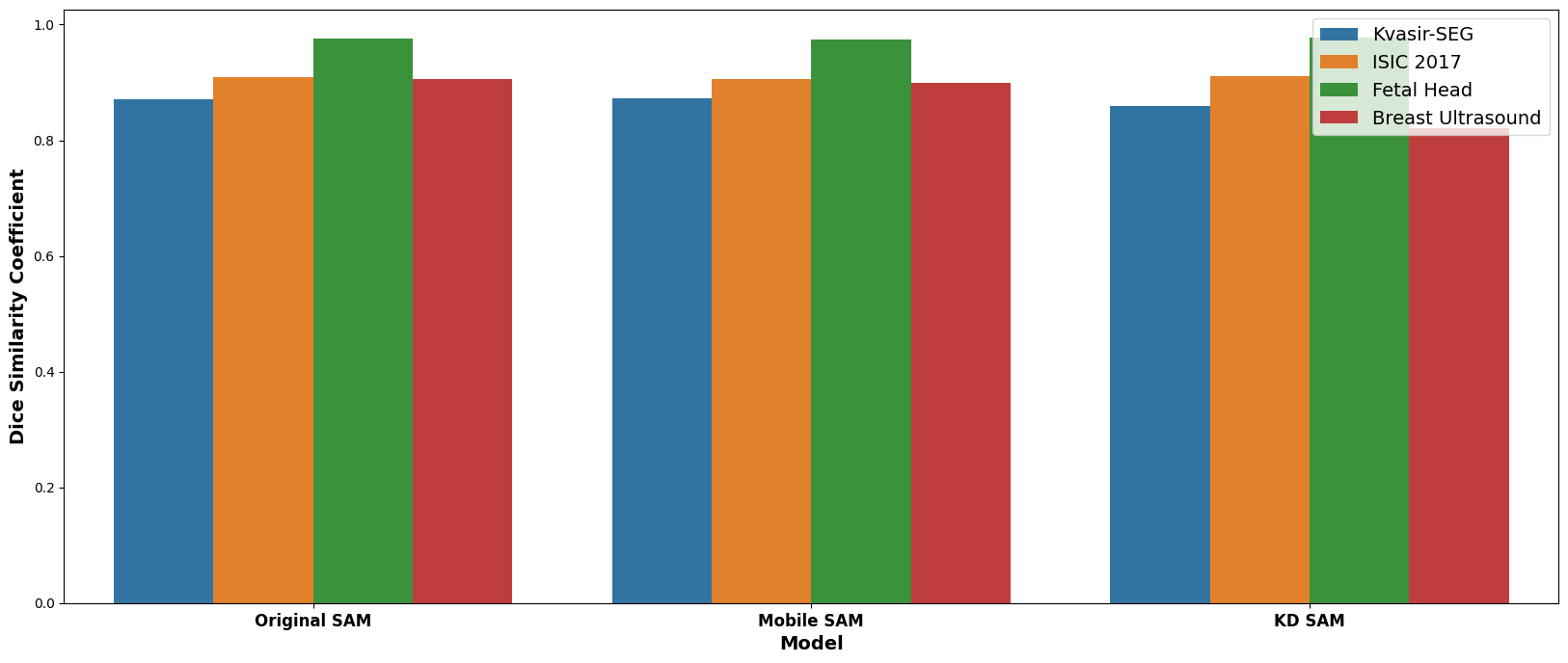}}
    %  \vspace{2.0cm}
    \centerline{(b) Bar Chart}\medskip
  \end{minipage}

  \caption{Comparative Statistical Analysis of Dice Coefficient Scores for Different
    Segmentation Models Across Various Datasets.}
  \label{fig:res}
\end{figure}

In challenging cases, such as detecting small polyps in the Kvasir-SEG dataset or accurately delineating melanoma boundaries in the ISIC 2017 dataset, the KD SAM model demonstrates a superior ability to capture fine details and produce smooth boundaries. This qualitative performance shown in Figure 3 supports the quantitative results, highlighting the model efficiency in medical image segmentation tasks.Overall, the separate test dataset results indicate that KD SAM effectively balances computational efficiency with segmentation accuracy, making it particularly suitable for medical image segmentation tasks in resource-constrained environments.

\begin{figure}[htb]

  \begin{minipage}[b]{1.0\linewidth}
    \centering
    \centerline{\includegraphics[width=8.5cm]{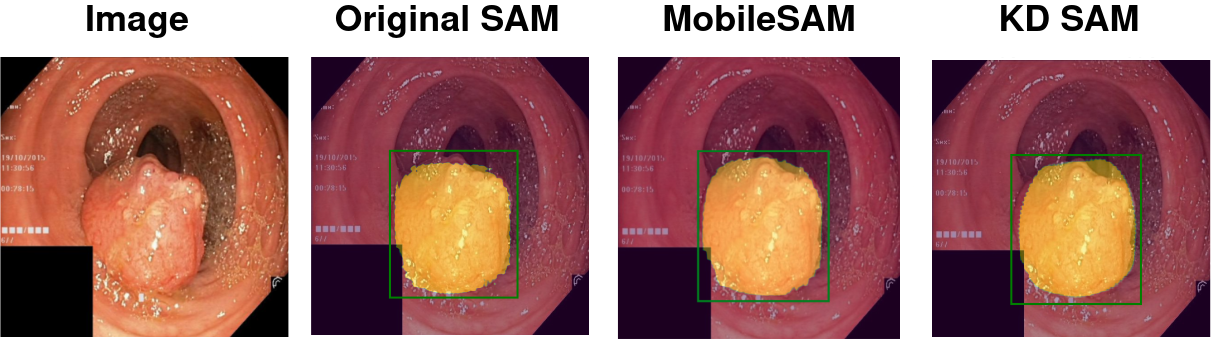}}
    %  \vspace{2.0cm}
    \centerline{(a) Kvasir-SEG}\medskip
  \end{minipage}
  \begin{minipage}[b]{1.0\linewidth}
    \centering
    \centerline{\includegraphics[width=8.5cm]{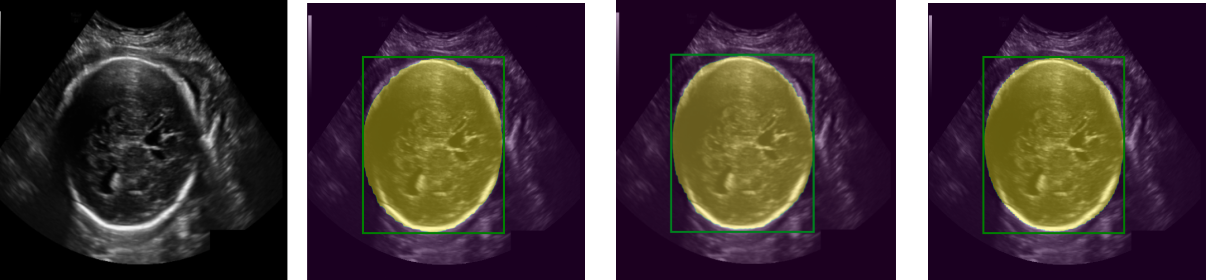}}
    %  \vspace{2.0cm}
    \centerline{(b) Fetal Head }\medskip
  \end{minipage}

  \begin{minipage}[b]{1.0\linewidth}
    \centering
    \centerline{\includegraphics[width=8.5cm]{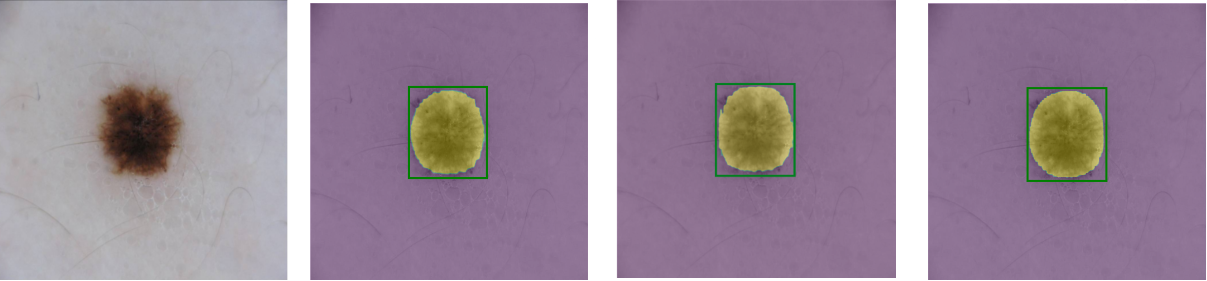}}
    %  \vspace{2.0cm}
    \centerline{(c) ISIC 2017 }\medskip
  \end{minipage}

  \begin{minipage}[b]{1.0\linewidth}
    \centering
    \centerline{\includegraphics[width=8.5cm]{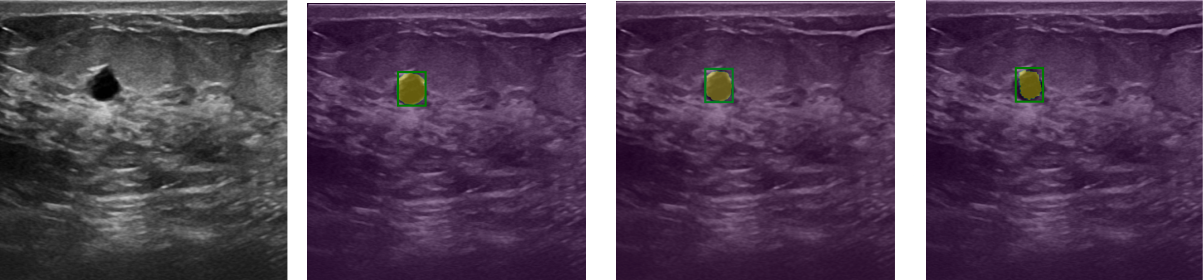}}
    %  \vspace{2.0cm}
    \centerline{(d) Breast Ultrasound}\medskip
  \end{minipage}
  \caption{Comparison of Segmentation Results against SAM and MobileSAM}
  \label{fig:res}
\end{figure}

\bibliographystyle{IEEEbib}
\bibliography{strings,refs}

\begin{thebibliography}{10}

\bibitem{kirillov2023segment}
Alexander Kirillov, Eric Mintun, Nikhila Ravi, Hanzi Mao, Chloe Rolland, Laura Gustafson, Tete Xiao, Spencer Whitehead, Alexander~C Berg, Wan-Yen Lo, et~al.,
\newblock ``Segment anything,''
\newblock in {\em Proceedings of the IEEE/CVF International Conference on Computer Vision}, 2023, pp. 4015--4026.

\bibitem{dosovitskiy2020image}
Alexey Dosovitskiy,
\newblock ``An image is worth 16x16 words: Transformers for image recognition at scale,''
\newblock {\em arXiv preprint arXiv:2010.11929}, 2020.

\bibitem{zhang2023faster}
Chaoning Zhang, Dongshen Han, Yu~Qiao, Jung~Uk Kim, Sung-Ho Bae, Seungkyu Lee, and Choong~Seon Hong,
\newblock ``Faster segment anything: Towards lightweight sam for mobile applications,''
\newblock {\em arXiv preprint arXiv:2306.14289}, 2023.

\bibitem{johnson2016perceptual}
Justin Johnson, Alexandre Alahi, and Li~Fei-Fei,
\newblock ``Perceptual losses for real-time style transfer and super-resolution,''
\newblock in {\em Computer Vision--ECCV 2016: 14th European Conference, Amsterdam, The Netherlands, October 11-14, 2016, Proceedings, Part II 14}. Springer, 2016, pp. 694--711.

\bibitem{guan2020differentiable}
Yushuo Guan, Pengyu Zhao, Bingxuan Wang, Yuanxing Zhang, Cong Yao, Kaigui Bian, and Jian Tang,
\newblock ``Differentiable feature aggregation search for knowledge distillation,''
\newblock in {\em Computer Vision--ECCV 2020: 16th European Conference, Glasgow, UK, August 23--28, 2020, Proceedings, Part XVII 16}. Springer, 2020, pp. 469--484.

\bibitem{liu2019structured}
Yifan Liu, Ke~Chen, Chris Liu, Zengchang Qin, Zhenbo Luo, and Jingdong Wang,
\newblock ``Structured knowledge distillation for semantic segmentation,''
\newblock in {\em Proceedings of the IEEE/CVF conference on computer vision and pattern recognition}, 2019, pp. 2604--2613.

\bibitem{chen2017learning}
Guobin Chen, Wongun Choi, Xiang Yu, Tony Han, and Manmohan Chandraker,
\newblock ``Learning efficient object detection models with knowledge distillation,''
\newblock {\em Advances in neural information processing systems}, vol. 30, 2017.

\bibitem{zhao2023fast}
Xu~Zhao, Wenchao Ding, Yongqi An, Yinglong Du, Tao Yu, Min Li, Ming Tang, and Jinqiao Wang,
\newblock ``Fast segment anything,''
\newblock {\em arXiv preprint arXiv:2306.12156}, 2023.

\bibitem{bolya2019yolact}
Daniel Bolya, Chong Zhou, Fanyi Xiao, and Yong~Jae Lee,
\newblock ``Yolact: Real-time instance segmentation,''
\newblock in {\em Proceedings of the IEEE/CVF international conference on computer vision}, 2019, pp. 9157--9166.

\bibitem{xiong2024efficientsam}
Yunyang Xiong, Bala Varadarajan, Lemeng Wu, Xiaoyu Xiang, Fanyi Xiao, Chenchen Zhu, Xiaoliang Dai, Dilin Wang, Fei Sun, Forrest Iandola, et~al.,
\newblock ``Efficientsam: Leveraged masked image pretraining for efficient segment anything,''
\newblock in {\em Proceedings of the IEEE/CVF Conference on Computer Vision and Pattern Recognition}, 2024, pp. 16111--16121.

\bibitem{zhou2023edgesam}
Chong Zhou, Xiangtai Li, Chen~Change Loy, and Bo~Dai,
\newblock ``Edgesam: Prompt-in-the-loop distillation for on-device deployment of sam,''
\newblock {\em arXiv preprint arXiv:2312.06660}, 2023.

\bibitem{he2016deep}
Kaiming He, Xiangyu Zhang, Shaoqing Ren, and Jian Sun,
\newblock ``Deep residual learning for image recognition,''
\newblock in {\em Proceedings of the IEEE conference on computer vision and pattern recognition}, 2016, pp. 770--778.

\bibitem{simonyan2014very}
Karen Simonyan and Andrew Zisserman,
\newblock ``Very deep convolutional networks for large-scale image recognition,''
\newblock {\em arXiv preprint arXiv:1409.1556}, 2014.

\bibitem{jha2020kvasir}
Debesh Jha, Pia~H Smedsrud, Michael~A Riegler, P{\aa}l Halvorsen, Thomas De~Lange, Dag Johansen, and H{\aa}vard~D Johansen,
\newblock ``Kvasir-seg: A segmented polyp dataset,''
\newblock in {\em MultiMedia modeling: 26th international conference, MMM 2020, Daejeon, South Korea, January 5--8, 2020, proceedings, part II 26}. Springer, 2020, pp. 451--462.

\bibitem{cassidy2022analysis}
Bill Cassidy, Connah Kendrick, Andrzej Brodzicki, Joanna Jaworek-Korjakowska, and Moi~Hoon Yap,
\newblock ``Analysis of the isic image datasets: Usage, benchmarks and recommendations,''
\newblock {\em Medical image analysis}, vol. 75, pp. 102305, 2022.

\bibitem{van2018automated}
Thomas~LA van~den Heuvel, Dagmar de~Bruijn, Chris~L de~Korte, and Bram~van Ginneken,
\newblock ``Automated measurement of fetal head circumference using 2d ultrasound images,''
\newblock {\em PloS one}, vol. 13, no. 8, pp. e0200412, 2018.

\bibitem{al2020dataset}
Walid Al-Dhabyani, Mohammed Gomaa, Hussien Khaled, and Aly Fahmy,
\newblock ``Dataset of breast ultrasound images,''
\newblock {\em Data in brief}, vol. 28, pp. 104863, 2020.

\bibitem{kingma2014adam}
Diederik~P Kingma,
\newblock ``Adam: A method for stochastic optimization,''
\newblock {\em arXiv preprint arXiv:1412.6980}, 2014.

\end{thebibliography}

\end{document}